\RequirePackage{lineno} 
\documentclass[twocolumn,aps,prl,superscriptaddress,showpacs,floatfix,longbibliography]{revtex4-1}
\usepackage{url}
\usepackage{cancel}
\usepackage[colorlinks,linkcolor=blue,citecolor=blue,filecolor=black,urlcolor=blue]{hyperref}
\usepackage{epsfig,graphics}
\usepackage{graphicx}% Include figure files
\usepackage{dcolumn}% Align table columns on decimal point
\usepackage{bm}% bold math
\usepackage[usenames]{color}
\usepackage{amssymb}
\usepackage{amsmath}
\usepackage{multirow}
\usepackage{float}
\usepackage{harpoon}
\usepackage{MnSymbol}
\usepackage{appendix}
\newcommand{\nch} {N_{\mathrm{ch}}}
\newcommand{\snn}{\mbox{$\sqrt{s_{\mathrm{NN}}}$}}
\newcommand{\pT} {p_{\mathrm{T}}}
\newcommand{\lr}[1]{\left\langle #1\right\rangle}
\newcommand{\Deta}{\Delta \eta}

\hfuzz=\maxdimen
\graphicspath{{figs2/}{figs1/}{./}}
\begin{document}
\title{Long- and short-range longitudinal flow decorrelations in the quark-gluon plasma}
\newcommand{\sbu}{Department of Chemistry, Stony Brook University, Stony Brook, NY 11794, USA}
\newcommand{\moe}{Key Laboratory of Nuclear Physics and Ion-beam Application (MOE), and Institute of Modern Physics, Fudan University, Shanghai 200433, China}
\newcommand{\fdu}{Shanghai Research Center for Theoretical Nuclear Physics, NSFC and Fudan University, Shanghai 200438, China}
\author{Jiangyong Jia}\email[Correspond to\ ]{jiangyong.jia@stonybrook.edu}\affiliation{\sbu}
\author{Shengli Huang}\affiliation{\sbu}
\author{Chunjian Zhang}\email[Correspond to\ ]{chunjianzhang@fudan.edu.cn}\affiliation{\fdu}\affiliation{\moe}\affiliation{\sbu}
\author{Somadutta Bhatta}\affiliation{\sbu}
 
\begin{abstract}
The longitudinal structure of the quark-gluon plasma (QGP) remains poorly constrained. Using a transport-model calculation of two-particle elliptic-flow correlations $V_{2\Delta}(\Delta\eta)$ in relative pseudorapidity $\Delta\eta$, we find two distinct initial-state components of flow decorrelation: a long-range component driven by the global geometry of participating nucleons and a short-range component reflecting local energy-deposition fluctuations. This short-range component extends to $|\Delta\eta|\sim 4$ and can be misidentified as final-state nonflow. For isobar-like nuclei with different shapes, the calculation predicts an experimentally testable fingerprint of this component: it dilutes the deformation contribution at small $|\Delta\eta|$, so that the $V_{2\Delta}(\Delta\eta)$ ratio between deformed and near-spherical systems, such as U+U-to-Au+Au, rises with $|\Delta\eta|$. Separating long- and short-range flow decorrelations from nonflow gives access to the longitudinal initial structure and dynamics of the QGP.
\end{abstract}
\maketitle

{\bf Introduction.} 
High-energy nuclear collisions serve as laboratories for generating and examining the quark-gluon plasma (QGP) under varied conditions~\cite{Busza:2018rrf}. Understanding the dynamics and properties of this exotic state requires precise control of its initial conditions (IC)~\cite{Berges:2020fwq}, including the distributions of colliding nucleons, their constituent partons, energy deposition patterns, and the 3D geometric profile. Recent studies have shown that comparing collision systems with similar masses but distinct nuclear shapes (isobar-like pairs) enables controlled variation of the IC~\cite{Giacalone:2021uhj,Jia:2021oyt,Nijs:2021kvn,Giacalone:2023hwk,Giacalone:2024luz,STAR:2024eky,STAR:2025rot}. Because common final-state-response uncertainties largely cancel in system ratios, this approach constrains the geometrical properties of the IC~\cite{Jia:2022ozr}.

\begin{figure}[!b]
\vspace*{-0.3cm}
\centering
\includegraphics[width=1\linewidth]{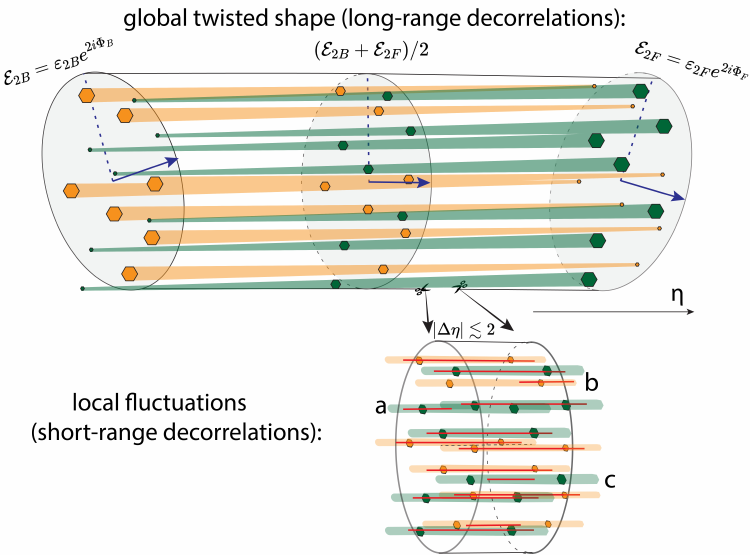}
\vspace*{-0.5cm}
\caption{\label{fig:1} Global and local sources of longitudinal shape fluctuations. Top: forward-going (green) and backward-going (orange) participants each sweep out a light-colored tube in which energy is deposited. The varying tube widths represent asymmetric longitudinal energy deposition, i.e., forward-going nucleons deposit more energy at positive $\eta$ and vice versa, so that the eccentricity interpolates gradually between $\mathcal{E}_{2,\mathrm{B}}$ and $\mathcal{E}_{2,\mathrm{F}}$. Bottom: deposited energy from each nucleon spreads over segments (red) spanning $|\Delta\eta|\lesssim1$--$2$. Within this range, most nucleons deposit energy coherently across nearby slices, but a few may deposit more energy in the left slice (a), the right slice (b), or between the two (c). This stochastic behavior leads to short-range decorrelations of $\mathcal{E}_{2}$ and $V_2$.}
\end{figure}

Most studies have focused on the transverse $xy$ profile of the IC near mid-pseudorapidity ($\eta\approx0$). However, the energy deposition process is inherently non-boost-invariant, so the IC fluctuate along $\eta$, giving rise to longitudinal fluctuations or decorrelations~\cite{Bozek:2015bna,Pang:2015zrq,Li:2019eni,Franco:2019ihq}. Consequently, the initial shape and transverse expansion of the QGP vary with $\eta$ even within a single collision event. In QCD-inspired string models (PYTHIA~\cite{Sjostrand:2019zhc}, HIJING~\cite{Gyulassy:1994ew}, AMPT~\cite{Lin:2004en}), nucleons deposit energy in color flux tubes whose endpoints in $\eta$ vary from tube to tube~\cite{Bozek:2015bna,Pang:2015zrq}. This produces both a global geometry that varies smoothly with $\eta$ and local stochastic fluctuations (Fig.~\ref{fig:1}). These longitudinal features manifest as rapidity decorrelations in bulk observables, including charged-particle multiplicity~\cite{PHOBOS:2006mfc,STAR:2009goo,ATLAS:2016rbh,ALICE:2017mtc}, anisotropic flow~\cite{CMS:2015xmx,ATLAS:2017rij,ALICE:2023tvh}, and average transverse momentum~\cite{ATLAS:2025ztg}.

We illustrate these decorrelations with elliptic flow ($V_2=v_2e^{2i\Psi}$), which characterizes the azimuthal distribution of particles in the $xy$-plane. $V_2$ captures the hydrodynamic response to the elliptic overlap region, quantified by the eccentricity vector $\mathcal{E}_2 (\eta)$. The two quantities follow an approximately linear relation $V_2 (\eta) \propto \mathcal{E}_2 (\eta)$~\cite{Niemi:2012aj}. The eccentricity $\mathcal{E}_2$ is defined from the transverse energy density $e(r,\phi)$ at a given $\eta$ as
\small\begin{align}\label{eq:1}
\mathcal{E}_2 = \varepsilon_2 e^{2i\Phi}\equiv -\frac{\int  e(r,\phi)  r^2 e^{2i\phi} rdrd\phi}{\int  e(r,\phi) r^2 rdrd\phi}\;.
\end{align}\normalsize
Asymmetric energy deposition produces an eccentricity $\mathcal{E}_2(\eta)$ that interpolates between the eccentricities of the forward- and backward-going participants, $\mathcal{E}_{2,\mathrm{F}}$ and $\mathcal{E}_{2,\mathrm{B}}$~\cite{Jia:2014ysa}. The initially produced QGP therefore has an elliptic shape that twists and deforms from $\mathcal{E}_{2,\mathrm{F}}$ to $\mathcal{E}_{2,\mathrm{B}}$ along the longitudinal direction, driving a final-state $V_2$ with similar $\eta$ dependence (top of Fig.~\ref{fig:1}). Local energy-deposition fluctuations instead add stochastic variations to $\mathcal{E}_2(\eta)$, and hence to $V_2(\eta)$, over shorter $\eta$ intervals (bottom of Fig.~\ref{fig:1}).

Experimentally, elliptic flow is accessed using the two-particle correlation (2PC) method, which measures its event-averaged two-point correlation:
\begin{align}\label{eq:2}
V_{2\Delta}(\eta_1,\eta_2)=\lr{V_2(\eta_1)V_2^*(\eta_2)}\propto \lr{\mathcal{E}_2(\eta_1)\mathcal{E}_2^*(\eta_2)}\;.
\end{align} 
Initial-state longitudinal fluctuations cause $V_{2\Delta}$ to decrease with increasing pair separation $\Delta\eta = \eta_1-\eta_2$. We refer to the component of $V_{2\Delta}$ that varies gradually with $\Delta\eta$, driven by the global participant geometry including nuclear deformation, as ``long-range'' correlations (LRC), and to the component that falls off over shorter $|\Delta\eta|$, driven by local energy-deposition fluctuations, as ``short-range'' correlations (SRC)~\cite{Bozek:2015bna}.

Measured two-particle correlations show a strong, centrality-dependent broadening of the near-side peak around $\Delta\eta=0$~\cite{STAR:2009ngv,ALICE:2016mfw}. This broadening resembles the SRC and has been attributed to longitudinal broadening of local fluctuations~\cite{Romatschke:2006bb,Gavin:2008ev}. Nonflow correlations from jet fragmentation and resonance decays, however, are unrelated to the initial geometry and also contribute significantly at small $|\Delta\eta|$. In 3D dynamical models the SRC extends to larger $|\Delta\eta|$ than nonflow, but the two remain difficult to separate experimentally.

In this Letter, we compare $V_{2\Delta}(\eta_1,\eta_2)$ with the flow projected onto global and local estimates of the initial eccentricity to separate initial-state-driven LRC and SRC from nonflow. By comparing U+U with Au+Au collisions, we identify a fingerprint of initial-state-driven SRC: a dilution of the nuclear deformation effect that leads to an increase of the ratio $V_{2\Delta,\mathrm{UU}}/V_{2\Delta,\mathrm{AuAu}}$ with $|\Delta\eta|$.  Combining the projections with this dependence, we estimate the long-range, short-range, and final-state nonflow contributions within the model.

{\bf Simulation setup.} 
We employ the AMPT transport model~\cite{Lin:2004en}, which simulates the full space-time evolution of heavy-ion collisions on an event-by-event basis. Nucleons are sampled from a deformed Woods-Saxon distribution, $\rho(r,\theta,\phi) \propto \left[1+\exp\left(\left(r-R_0\left(1+\beta Y_{2,0}(\theta)\right)\right)/a_0\right)\right]^{-1}$, where $\beta$ is the quadrupole deformation parameter, $R_0$ is the nuclear radius, and $a_0$ is the skin thickness. Three collision systems are simulated with geometrical parameters listed in Table~\ref{tab:1}. Collision events are generated at $\snn=193$ GeV for U+U and $\snn=200$ GeV for Au+Au using the string-melting mode with a partonic cross-section of 3~mb. With this configuration, AMPT reproduces the measured rapidity distribution of particle production~\cite{Lin:2004en} and isobar ratios of $v_2$ and $v_3$~\cite{Jia:2021oyt}; AMPT-generated initial conditions also describe measured flow decorrelations after hydrodynamic evolution~\cite{Pang:2015zrq,Wu:2018cpc}. Isobar ratios of $v_2$ and $v_3$ in AMPT are also insensitive to the partonic cross section (1.5--10~mb), hadronization, and hadronic cascade~\cite{Zhang:2022fou}, so the system ratios studied here are expected to be robust against these final-state choices. Final-state particles are selected in the transverse momentum range $0.2<\pT<3$ GeV/$c$. Event centrality is defined by the charged-particle multiplicity $\nch$ within $|\eta|<0.5$. We focus on the 0--5\% most central collisions, where nuclear-deformation effects are most prominent.

{\bf Analysis method.} 
We calculate three related eccentricities: the rapidity-independent participant-nucleon eccentricity $\mathcal{E}_{2,\mathrm{N}}$, which characterizes the global geometry (End Matter); the local parton eccentricity $\mathcal{E}_{2,\mathrm{q}}(\eta)$, Eq.~\eqref{eq:1} evaluated from the parton energy density at each $\eta$; and the reference eccentricity $\mathcal{E}_{2,\mathrm{q}0}\equiv\mathcal{E}_{2,\mathrm{q}}(|\eta|<0.4)$, calculated from the parton energy density integrated over $|\eta|<0.4$. All three are evaluated before partonic transport. The narrow window ties the reference to the local fluctuations at midrapidity while retaining adequate statistical precision; changing it to $|\eta|<0.2$ or $|\eta|<0.6$ gives consistent results.

\begin{table}[!h]
\vspace*{-0.3cm}
\centering
\caption{\label{tab:1} Nuclear-density parameters used for the three collision systems:  quadrupole deformation $\beta$, radius $R_0$, and skin thickness $a_0$~\cite{Ryssens:2023fkv}.}
\small{\begin{tabular}{c|c|c|c}\hline
  System    &$^{238}$U+$^{238}$U  & $^{238}$U+$^{238}$U & $^{197}$Au+$^{197}$Au\\\hline
  $\beta$   & 0.28 & 0.0 & $-0.14$ \\
  $R_0$(fm) & 6.81 & 6.81 & 6.62 \\
  $a_0$(fm) & 0.55 & 0.55 & 0.52 \\\hline
\end{tabular}}\normalsize
\end{table}

We characterize the flow in two ways and compare them to separate the decorrelation components. First, we calculate the 2PC elliptic flow coefficient $V_{2\Delta}(\eta_1,\eta_2)$ within the typical detector coverage $|\eta_1|,|\eta_2|<2$ at RHIC and the LHC.  The 1D $V_{2\Delta}(\Delta\eta)$ is obtained by averaging over pairs at fixed $\Delta\eta$ within this acceptance. It extends to $|\Delta\eta|=4$ and contains initial-state-driven LRC and SRC, as illustrated in Fig.~\ref{fig:1}, together with final-state nonflow contributions.

\begin{figure*}[htbp]
\centering
\vspace*{-0.4cm}
\includegraphics[width=0.75\linewidth]{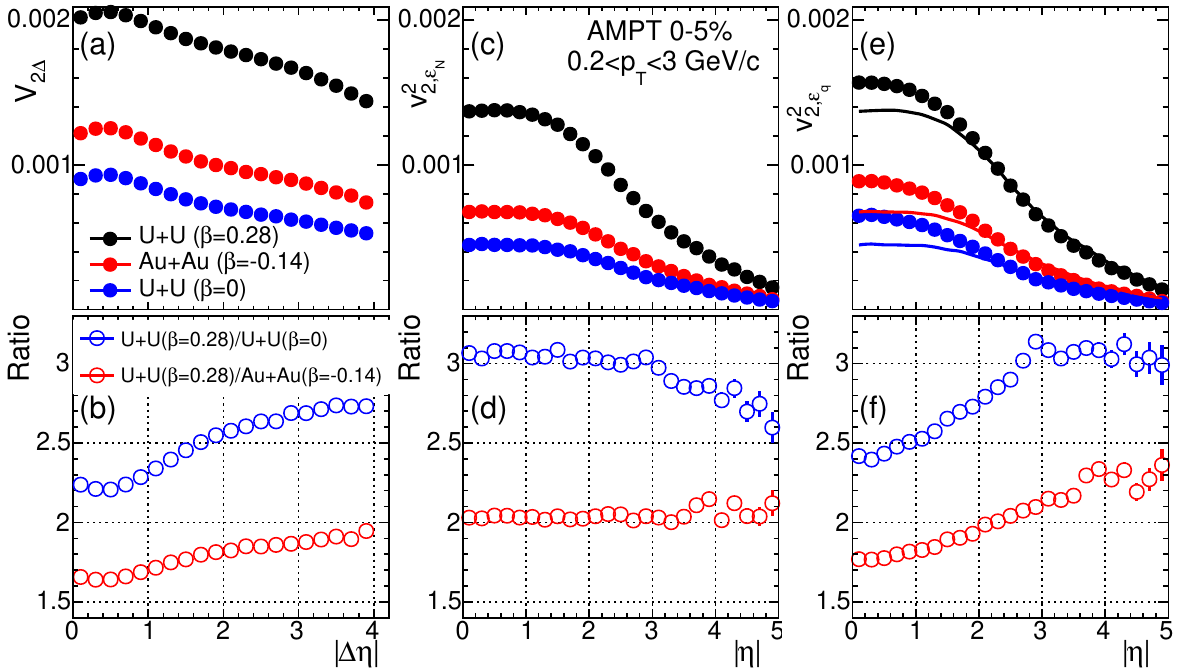}
\vspace*{-0.3cm}
\caption{\label{fig:2} Longitudinal dependence of two-particle correlations and flow--eccentricity projections. Elliptic flow from the 2PC method versus $\Delta\eta$ (a), and the squared flow projected onto $\mathcal{E}_{2,\mathrm{N}}$ (c) and onto the fixed reference $\mathcal{E}_{2,\mathrm{q}0}$ (e) versus $\eta$, for the three systems listed in Table~\ref{tab:1}. Bottom panels (b), (d), and (f) show the corresponding ratios between U+U with and without deformation (open blue circles) and between deformed U+U and Au+Au (open red circles). The curves from panel (c) are overlaid in panel (e) for direct comparison of $v_{2,\varepsilon_{\mathrm{q}}}^2$ with $v_{2,\varepsilon_{\mathrm{N}}}^2$.}
\vspace*{-0.2cm}
\end{figure*}

Second, we calculate the flow projected onto the eccentricity vectors:
\small{\begin{align}\label{eq:3}
v_{2,\varepsilon_{\mathrm{N}}}(\eta) \equiv\frac{\lr{V_2(\eta) \mathcal{E}_{2,\mathrm{N}}^*}}{\sqrt{\lr{|\mathcal{E}_{2,\mathrm{N}}|^2}}}\;,\;v_{2,\varepsilon_{\mathrm{q}}}(\eta) \equiv\frac{\lr{V_2(\eta) \mathcal{E}_{2,{\mathrm{q}0}}^*}}{\sqrt{\lr{|\mathcal{E}_{2,{\mathrm{q}0}}|^2}}}\;
\end{align}}\normalsize
where $\lr{}$ denotes averaging over events in a specific centrality class. Equation~\eqref{eq:3} is a projection in the statistical sense: it equals the Pearson correlation coefficient $Q(\mathcal{E}_2,V_2)$ multiplied by $\sqrt{\lr{|V_2|^2}}=V_{2\Delta}(\eta,\eta)^{1/2}$~\cite{Giacalone:2020byk} (End Matter). The two quantities retain the part of the final-state flow correlated with the global geometry and with the fixed midrapidity parton geometry, respectively; the latter also carries local energy-deposition fluctuations. Because both references are fixed before final-state evolution, nonflow correlations drop out of the projections.

We isolate the deformation-induced long-range contribution as~\cite{Jia:2021oyt}
\small{\begin{align}\label{eq:4}
 V_{2\Delta,\beta}(\Deta) = V_{2\Delta}\{\beta=0.28\}-V_{2\Delta}\{\beta=0\}\approx b(\Deta)\beta^2\;,
\end{align}}\normalsize
and write the total 2PC correlation schematically as
\begin{align}\label{eq:5}
V_{2\Delta}(\Deta) = a(\Deta)+b(\Deta) \beta^2 + \mathrm{src}(\Deta) + \mathrm{nf}(\Deta)\;.
\end{align}
Here, $a(\Delta\eta)$ and $\mathrm{src}(\Delta\eta)$ are the initial-state-driven LRC and SRC in collisions of spherical nuclei, $b(\Delta\eta) \beta^2$ is the nuclear deformation contribution, whose long-range structure is similar to that of $a(\Delta\eta)$~\cite{Zhang:2024bcb}, and $\mathrm{nf}(\Delta\eta)$ is final-state nonflow.

\begin{figure*}[htbp]
\centering
\vspace*{-0.4cm}
\includegraphics[width=0.75\linewidth]{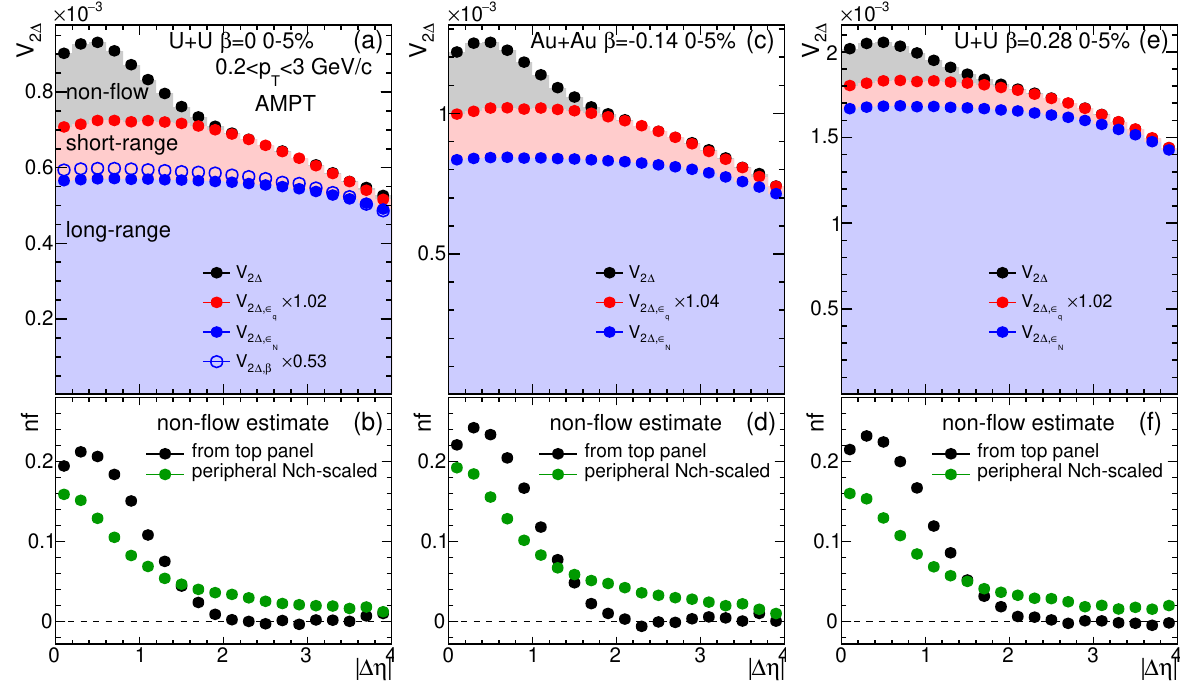}
\vspace*{-0.3cm}
\caption{\label{fig:3} Decomposition of two-particle elliptic-flow correlations into long-range, short-range, and nonflow components (shaded areas). Top panels compare $V_{2\Delta}$ (solid black), $V_{2\Delta,\varepsilon_{\mathrm{q}}}$ (solid red), and $V_{2\Delta,\varepsilon_{\mathrm{N}}}$ (solid blue) in 0--5\% central collisions of (a) undeformed U+U, (c) Au+Au, and (e) deformed U+U. The red curves are normalized to the large-$|\Delta\eta|$ tails of $V_{2\Delta}$, with the applied factors shown in the panels. Their size, 1.02--1.04, quantifies the part of the flow at large $\Delta\eta$ not captured by the projection. In panel (a), deformation-induced flow $V_{2\Delta,\beta}$ is scaled to match the tail of $V_{2\Delta,\varepsilon_{\mathrm{N}}}$ (open blue circles). Panels (b), (d), and (f) compare the estimated nonflow (black) with the peripheral-scaling estimate obtained from 90--100\% peripheral collisions selected by $N_{\mathrm{part}}$ (green).}
\end{figure*}

To separate these components, we bring the projections of Eq.~\eqref{eq:3} onto the same $\Delta\eta$ footing as $V_{2\Delta}$ by convolving $v_{2,\varepsilon}(\eta)$ with itself over the region where both $|\eta|<2$ and $|\eta-\Deta|<2$, with appropriate normalization:
\begin{align}\label{eq:6}
V_{2\Delta,\varepsilon}(\Deta) = \left.\int v_{2,\varepsilon}(\eta)\,v_{2,\varepsilon}(\eta-\Deta)\,d\eta \right/\!\int d\eta \;,
\end{align}
where $\varepsilon$ stands for either $\varepsilon_{\mathrm{N}}$ or $\varepsilon_{\mathrm{q}}$. Since $V_{2\Delta,\varepsilon_{\mathrm{N}}}$ retains only the long-range terms of Eq.~\eqref{eq:5} while $V_{2\Delta,\varepsilon_{\mathrm{q}}}$ also retains the local fluctuations, we identify the long-range component with $V_{2\Delta,\varepsilon_{\mathrm{N}}}$, the short-range component with the difference $V_{2\Delta,\varepsilon_{\mathrm{q}}}-V_{2\Delta,\varepsilon_{\mathrm{N}}}$, and the residual $V_{2\Delta}-V_{2\Delta,\varepsilon_{\mathrm{q}}}$ primarily with nonflow. The performance of this decomposition is examined in Fig.~\ref{fig:3}.

Nuclear deformation offers a way to expose the short-range component experimentally. To see this, we examine the ratio between U+U collisions with and without deformation:
\small{\begin{align}\nonumber
R_{V_{2\Delta}}(\Deta) &= \frac{a(\Deta)+b(\Deta) \beta^2 + \mathrm{src}(\Deta) + \mathrm{nf}(\Deta)}{a(\Deta)+ \mathrm{src}(\Deta) + \mathrm{nf}(\Deta)}\\\label{eq:7}
&=1+ c\beta^2\left/\left(1+ \frac{\mathrm{src}(\Deta)+\mathrm{nf}(\Deta)}{a(\Deta)}\right)\right.
\end{align}}\normalsize
where $c=b(\Delta\eta)/a(\Delta\eta)$ is approximately constant~\cite{Zhang:2024bcb}. According to Eq.~\eqref{eq:7}, both SRC and nonflow dilute $R_{V_{2\Delta}}$ over their respective $\Delta\eta$ ranges, so that $R_{V_{2\Delta}}$ rises with $|\Delta\eta|$ as each contribution dies out. In experiment, the analogous ratio $V_{2\Delta,\mathrm{UU}}/V_{2\Delta,\mathrm{AuAu}}$ is directly accessible. The moderate $\beta_{\mathrm{Au}}$ reduces the deformation contrast, but the ratio is expected to show the same qualitative dilution and rise with $|\Delta\eta|$. Their different $\Delta\eta$ ranges make this dilution a distinctive fingerprint of initial-state-driven SRC and final-state nonflow.

{\bf Flow projections onto initial geometry.} 
Figure~\ref{fig:2}(a) displays $V_{2\Delta}$ as a function of $\Delta\eta$ for all three collision systems. The distribution shows a peak at $|\Delta\eta|\approx 0.4$, a rapid decrease to $|\Delta\eta|\sim 1.5$, and a slower decrease beyond $|\Delta\eta|\sim 2$; we show below that the former is dominated by nonflow and the latter by initial-state-driven SRC.

Figure~\ref{fig:2}(b) presents ratios of $V_{2\Delta}$ between U+U with and without deformation, and between deformed U+U and Au+Au. The ratios rise with $|\Delta\eta|$, as expected from Eq.~\eqref{eq:7}. The rapid rise ends at $|\Delta\eta|\approx 1.5$ and a slower rise continues to the edge of the acceptance, consistent with $\mathrm{nf}(\Delta\eta)$ being largely confined to $|\Delta\eta|\lesssim 1.5$ and $\mathrm{src}(\Delta\eta)$ extending further.

Figures~\ref{fig:2}(c) and (e) show $v_{2,\varepsilon_{\mathrm{N}}}^2$ and $v_{2,\varepsilon_{\mathrm{q}}}^2$ as functions of $\eta$. These quantities are smaller than $V_{2\Delta}$ because they retain only the components correlated with the chosen initial geometry. The corresponding ratios between U+U with and without deformation appear in panels (d) and (f). The ratio $R_{v_{2,\varepsilon_{\mathrm{N}}}^2}$ is nearly independent of $\eta$, as expected because $v_{2,\varepsilon_{\mathrm{N}}}$ retains only the long-range terms $a$ and $b\beta^2$ of Eq.~\eqref{eq:5}. In contrast, $R_{v_{2,\varepsilon_{\mathrm{q}}}^2}$ is suppressed at small $|\eta|$, with an $\eta$ dependence closely resembling the $\Delta\eta$ dependence in Fig.~\ref{fig:2}(b): local fluctuations captured by the reference $\mathcal{E}_{2,{\mathrm{q}0}}$ generate additional flow at midrapidity.

This interpretation is reinforced in Fig.~\ref{fig:2}(e), where $v_{2,\varepsilon_{\mathrm{q}}}>v_{2,\varepsilon_{\mathrm{N}}}$ at $|\eta|<2$ but the two approach each other as larger $|\eta|$. The corresponding rise of $R_{v_{2,\varepsilon_{\mathrm{q}}}^2}$ in Fig.~\ref{fig:2}(f) is naturally accounted for by the initial-state short-range component. The weaker but similar U+U-to-Au+Au variations in the bottom panels reflect partial compensation from the moderate deformation of Au.

{\bf Decomposition of flow decorrelations.} Figure~\ref{fig:3}(a), (c), and (e) compare $V_{2\Delta,\varepsilon_{\mathrm{N}}}$ and $V_{2\Delta,\varepsilon_{\mathrm{q}}}$ of Eq.~\eqref{eq:6} with $V_{2\Delta}$. After a small (2--4\%) tail adjustment, $V_{2\Delta,\varepsilon_{\mathrm{q}}}$ reproduces the shape of $V_{2\Delta}$ over the full $|\Delta\eta|>2$ range, indicating that the projection onto $\mathcal{E}_{2,\mathrm{q}0}$ accounts for nearly all of $V_{2\Delta}$ at large $|\Delta\eta|$. The need for this adjustment may reflect additional initial-state-driven contributions to $V_{2\Delta}$ that are not captured by this projection, including subleading modes or nonlinear terms in the initial geometry~\cite{Gardim:2011xv,Sousa:2024msh}. By contrast, $V_{2\Delta,\varepsilon_{\mathrm{N}}}$ is much flatter and resembles the deformation-induced component $V_{2\Delta,\beta}$ of Eq.~\eqref{eq:4}. Their similar $\Delta\eta$ dependences support the assumption in Eq.~\eqref{eq:7} that $c=b/a$ varies weakly with $\Delta\eta$.

These comparisons support the decomposition defined after Eq.~\eqref{eq:6}, shown by the shaded areas in Fig.~\ref{fig:3} with the scaled $V_{2\Delta,\varepsilon_{\mathrm{q}}}$. The long-range component is the largest. The short-range component from local fluctuations is next; it begins to fall at $|\Delta\eta|>2$, reflecting their typical coherence length, but remains visible out to $|\Delta\eta|\sim4$. The residual is concentrated within $|\Delta\eta|<2$ and is attributed primarily to nonflow. The bottom panels, Fig.~\ref{fig:3}(b), (d), and (f), compare this projection residual (black) with a peripheral-scaling estimate (green), obtained by assuming that nonflow is unmodified by the medium and scales inversely with $\nch$~\cite{STAR:2024eky}: $\mathrm{nf}_s(\Delta\eta) =  V_{2\Delta,90\text{--}100\%} N_{\mathrm{ch},90\text{--}100\%}/N_{\mathrm{ch},0\text{--}5\%}$\footnote{90--100\% centrality events are selected based on the number of participating nucleons $N_{\mathrm{part}}$ rather than $\nch$ to avoid selection bias.}. The peripheral-scaling estimate is broader but has a 30--40\% smaller integrated magnitude than the projection residual over $|\Delta\eta|<2$. The away-side jet tail of the peripheral-scaling estimate is absent from the projection estimate because jet quenching in central AMPT suppresses the $\cos(2\Delta\phi)$ component of the away-side jet.

\begin{figure}[htbp]
\centering
\vspace*{-0.4cm}
\includegraphics[width=0.7\linewidth]{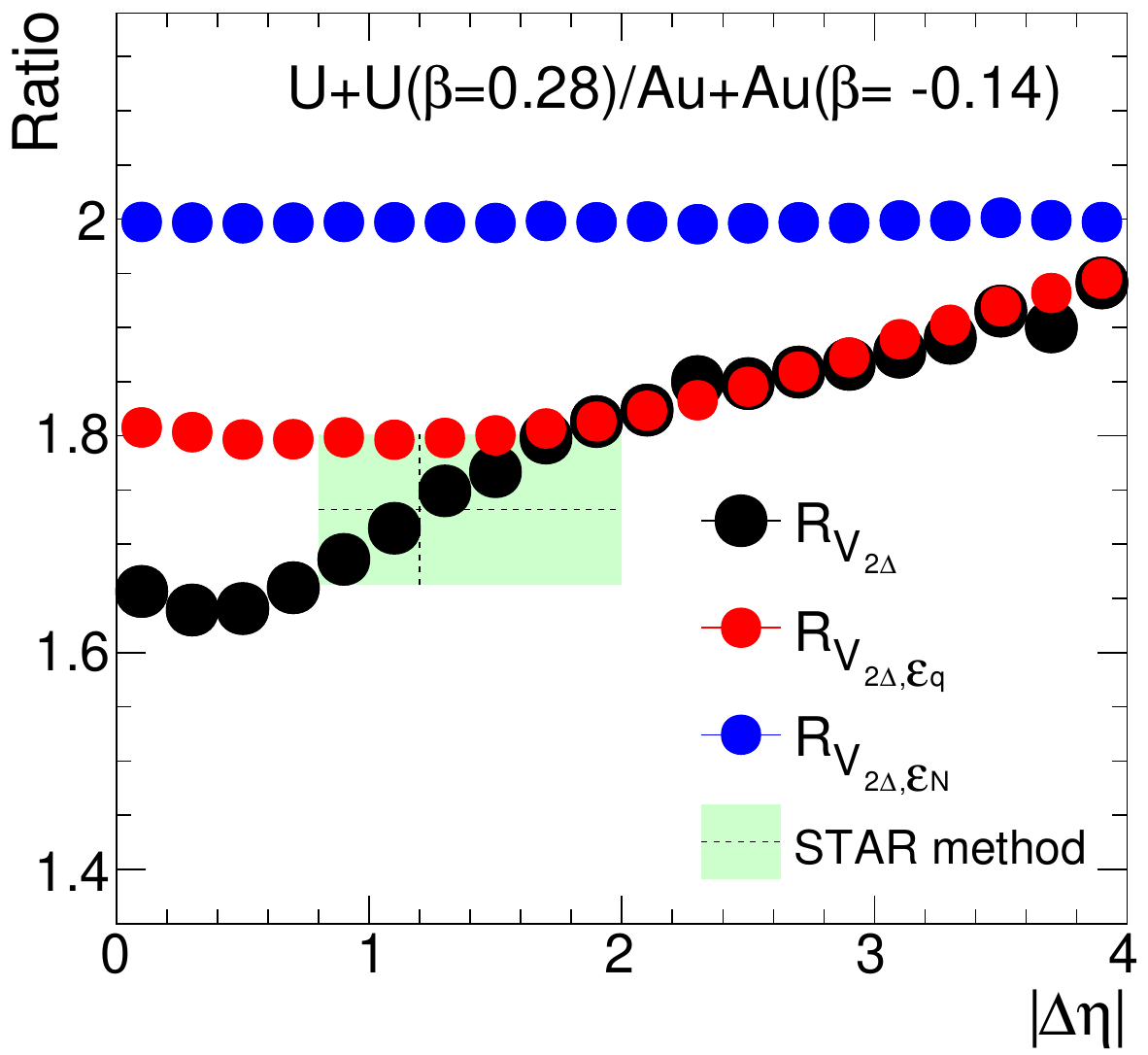}
\vspace*{-0.3cm}
\caption{\label{fig:4} Nuclear-shape ratios expose short-range longitudinal correlations. Ratios of $V_{2\Delta}$ (black), $V_{2\Delta,\varepsilon_{\mathrm{N}}}$ (blue, global correlations only), and $V_{2\Delta,\varepsilon_{\mathrm{q}}}$ (red, global plus short-range correlations) between deformed U+U and Au+Au collisions in 0--5\% centrality in AMPT. The green box marks the STAR two-subevent selection implemented in AMPT at an average $|\Delta\eta|=1.2$ and its quoted 4\% nonflow uncertainty~\cite{STAR:2024eky}. The rising trend, predicted by AMPT, is the experimentally testable fingerprint of initial-state-driven SRC and is expected to be generic for any isobar-like pair with significantly different $\beta$.}
\end{figure}

Figure~\ref{fig:4} presents the main prediction: within AMPT, the U+U-to-Au+Au ratio rises with $|\Delta\eta|$ because initial-state-driven SRC dilute the deformation signal over a wider range than nonflow. The figure compares the three ratios between deformed U+U and Au+Au collisions (the black points repeat the red circles of Fig.~\ref{fig:2}(b)). The ordering $R_{V_{2\Delta}}\lesssim R_{V_{2\Delta,\varepsilon_\mathrm{q}}}<R_{V_{2\Delta,\varepsilon_{\mathrm{N}}}}$ follows the logic of Eq.~\eqref{eq:7}: the gap between $R_{V_{2\Delta}}$ and $R_{V_{2\Delta,\varepsilon_\mathrm{q}}}$ is dilution by nonflow, and the gap between $R_{V_{2\Delta,\varepsilon_\mathrm{q}}}$ and $R_{V_{2\Delta,\varepsilon_{\mathrm{N}}}}$ is dilution by initial-state-driven SRC. Within the approximate decomposition, the nonflow component accounts for about 40\% of the suppression in $R_{V_{2\Delta}}$ at $|\Delta\eta|<2$, while the remaining suppression extends to $|\Delta\eta|\sim4$ and is associated with initial-state-driven SRC.

{\bf Impact on nuclear-shape extraction.} 
Recent studies have proposed using $R_{V_{2\Delta}}$ to probe the quadrupole deformation parameter $\beta$~\cite{Giacalone:2021udy,Magdy:2022cvt,ALICE:2024nqd}. In particular, STAR measured $R_{V_{2\Delta}}$ within $|\Delta\eta|<2$ using a two-subevent method and compared it with hydrodynamic model predictions to constrain uranium deformation~\cite{STAR:2024eky}. The $\beta_{\mathrm{U}}$ extracted from $R_{V_{2\Delta}}$ ($\beta_{\mathrm{U}}\sim0.24$) is smaller than values obtained from other observables ($\beta_{\mathrm{U}}\sim0.28$ from $p_{\mathrm{T}}$--$p_{\mathrm{T}}$ and $v_2^2$--$p_{\mathrm{T}}$ correlations). This discrepancy cannot be explained by residual nonflow contributions (End Matter).

The presence of initial-state-driven SRC extending to $|\Delta\eta|\sim4$ offers a plausible resolution to this puzzle, since such correlations dilute the deformation contribution and reduce $R_{V_{2\Delta}}$ (Eq.~\eqref{eq:7}). In the boost-invariant limit of the initial conditions used in the cited hydrodynamic comparisons, the ratio without nonflow is flat in $|\Delta\eta|$. In contrast, AMPT with longitudinally fluctuating initial conditions predicts the rising trend in Fig.~\ref{fig:4} for $\beta_{\mathrm{U}}=0.28$. This rising trend is therefore a distinctive signature of the longitudinal initial conditions rather than of the hydrodynamic evolution, and its measurement offers a direct test of the local-deposition picture.

{\bf Broader implications.} 
In flow analyses, the near-side enhancement and broadening of $V_{2\Delta}(\Delta\eta)$ have often been treated as nonflow. Within AMPT, most of this broadening instead arises from flow correlated with local initial-state geometry. The origin of this broadening in data, however, is not established. In addition to string-based energy deposition, as realized in AMPT, possible contributions include strong longitudinal expansion~\cite{Romatschke:2006bb,Gavin:2008ev} and thermal fluctuations~\cite{Kapusta:2011gt,Sakai:2020pjw,Yan:2025ath}.

Initial-state short-range correlations also affect the interpretation of flow decorrelations based on the factorization ratio $r_2(\eta) = \lr{V_2(-\eta)V_2^*(\eta_{\mathrm{ref}})}/\lr{V_2(\eta)V_2^*(\eta_{\mathrm{ref}})}$~\cite{CMS:2015xmx,ATLAS:2017rij}, where $\eta_{\mathrm{ref}}$ is typically chosen from the forward region (e.g., $\eta_{\mathrm{ref}}>4$) with at least a 2-unit gap from the measurement region (e.g., $|\eta|<2$). A finite-range SRC that does not reach $\eta_{\mathrm{ref}}$ leaves little trace in $r_2$, even though it still reduces the direct correlation between $V_2(-\eta)$ and $V_2(\eta)$; $r_2(\eta)$ therefore misses much of the short-range decorrelation.

Our conclusion concerns how the 3D initial condition is prepared rather than the detailed final-state dynamics. AMPT provides one realization of local fluctuating energy deposition. Other 3D initial-state frameworks deposit energy through strings~\cite{Shen:2017bsr} or color flux tubes~\cite{Schenke:2016ksl} and are expected to contain analogous forms of longitudinal locality. Measuring the $\Delta\eta$ dependence of the deformation signal would test this expectation.

{\bf Summary.} 
Within a transport-model calculation, we find that longitudinal flow decorrelations contain a slowly varying component correlated with the global participant geometry and a shorter-range component correlated with the local parton geometry. The local component extends to $|\Delta\eta|\sim4$ and is larger than the estimated nonflow contribution. Collisions of isobar-like nuclei with different shapes expose this component: it dilutes the deformation signal at small $|\Delta\eta|$, so that the ratio of $V_{2\Delta}$ between the two systems rises with $|\Delta\eta|$. Comparing the flow projected onto the participant and local parton eccentricities with $V_{2\Delta}$ separates these components from nonflow.

The impact of short-range correlations and nonflow was revealed here through the large nuclear deformation, and ratios between isobar-like systems with different nuclear shapes therefore provide sensitive probes of the QGP's rapidity structure. This method extends naturally to triangular flow decorrelations $V_{3\Delta}$, for example through $^{96}$Zr+$^{96}$Zr and $^{96}$Ru+$^{96}$Ru comparisons, where octupole deformation in $^{96}$Zr enhances $R_{V_{3\Delta}}$~\cite{STAR:2021mii,Zhang:2021kxj} analogously to Eq.~\eqref{eq:7}. Another promising direction is radial-flow decorrelations, recently measured at the LHC with two-particle correlation methods~\cite{ATLAS:2025ztg,ALICE:2025iud}. These possibilities can be pursued with existing large-acceptance LHC data and with dedicated isobar runs~\cite{Jia:2022ozr}.

J. Jia, S. Huang and S. Bhatta are supported by the U.S. Department of Energy, Office of Science, Office of Nuclear Physics, under DOE Awards No. DE-SC0024602. C. Zhang is supported by the National Key Research and Development Program of China under Contract No. 2024YFA1612600, Shanghai Pujiang Talents Program under Contract No. 24PJA009 and the NSFC under Contract No. 12547102. Claude and ChatGPT AIs were used to verify formulas and to improve the paper draft.

\bibliography{ref}{}

\section*{End Matter}
{\bf Relation of Eq.~\eqref{eq:3} to the flow--eccentricity correlator.} The Pearson correlation coefficient between the flow and an eccentricity vector is $Q(\mathcal{E}_2,V_2)=\lr{\mathcal{E}_2V_2^*}/\sqrt{\lr{|\mathcal{E}_2|^2}\lr{|V_2|^2}}$~\cite{Giacalone:2020byk}. Equation~\eqref{eq:3} restores the flow magnitude, $v_{2,\varepsilon}=Q(\mathcal{E}_2,V_2)\sqrt{\lr{|V_2|^2}}$, so that $v_{2,\varepsilon}$ can be compared directly with $V_{2\Delta}^{1/2}$. Equation~\eqref{eq:3} is thus a projection in a statistical sense, akin to principal-component analysis~\cite{Mazeliauskas:2015vea}, and applies separately for the reference $\mathcal{E}_{2,\mathrm{N}}$ or $\mathcal{E}_{2,\mathrm{q}0}$. To see what it represents, write $V_2=k\,\mathcal{E}_2+d$, where $k$ is a response coefficient and $d$ collects other initial-state-driven sources, satisfying $\lr{d\mathcal{E}_2^*}=0$. The imaginary part of $\lr{V_2\mathcal{E}_2^*}$ averages to zero by symmetry, so $\lr{V_2\mathcal{E}_2^*}=\mathrm{Re}\lr{V_2\mathcal{E}_2^*}=k\lr{|\mathcal{E}_2|^2}$ and Eq.~\eqref{eq:3} gives $k\sqrt{\lr{|\mathcal{E}_2|^2}}$, the part of the flow correlated with that reference.

{\bf Global nature of participant eccentricity.} A nucleon-level eccentricity has no counterpart in a narrow $\eta$ slice, because assigning a participant contribution to a given $\eta$ requires the longitudinal energy-deposition prescription encoded by $\mathcal{E}_{2,\mathrm{q}}(\eta)$. The comparison made here is therefore between the global participant geometry, $\mathcal{E}_{2,\mathrm{N}}$, and its local realization through energy deposition, rather than between two windows of different width.

{\bf The nonflow estimates.} We compare two estimates in Fig.~\ref{fig:3} rather than switching jet production on and off. In AMPT, the minijets contributing to these low-$\pT$ correlations are part of the underlying HIJING event and also affect the subsequent evolution, so disabling them would not isolate an additive nonflow component. Furthermore, part of the nonflow comes from resonance decays, which would not be reflected by such a switch.

{\bf Nonflow in the two-subevent selection.} To represent the STAR two-subevent procedure, we use the 2PC selection $0.8<|\Delta\eta|<2$. For this configuration, the residual fractions attributed primarily to nonflow are 6.8\% for Au+Au and 4.2\% for deformed U+U (Table~\ref{tab:2}). Correcting the ratio by these fractions changes it by $(1-0.042)/(1-0.068)-1=2.8\%$, within the nonflow uncertainty quoted by STAR~\cite{STAR:2024eky} and indicated by the green box in Fig.~\ref{fig:4}.

\begin{table}[!h]
\centering
\caption{\label{tab:2} Fractions of the integrated $V_{2\Delta}$ attributed to nonflow in 0--5\% centrality in AMPT for the full-event and two-subevent selections used in this calculation, obtained by integrating the black points in the bottom panels of Fig.~\ref{fig:3}.}
\small{\begin{tabular}{c|c|c}\hline
                        & full-event & two-subevent\\
                        & $|\Deta|<2$  & $0.8<|\Deta|<2$ \\\hline
 U+U $\beta=0.28$  & 6.9 \%  & 4.2 \%  \\
 U+U $\beta=0$     & 14.4 \% &  8.7 \% \\
Au+Au $\beta=-0.14$& 11.8 \%  & 6.8 \%\\\hline
\end{tabular}}\normalsize
\end{table}
\end{document}